# The large magnetocaloric effect and refrigerant capacity in nanocrystalline/amorphous $Gd_3Ni/Gd_{65}Ni_{35}$ composite microwires

Y.F. Wang[1,2], Y.Y. Yu[1], H. Belliveau[1], N.T.M. Duc[1,3], H.X. Shen[1,4,*], J. F. Sun[4], J.S. Liu[5], F.X. Qin[2,*], S.C. Yu[6], H. Srikanth[1], and M.H. Phan[1,*]

[1]Department of Physics, University of South Florida, Tampa, Florida 33620, USA

[2]Institute for Composites Science Innovation (InCSI), School of Materials Science and Engineering, Zhejiang University, 38 Zheda Road, Hangzhou, 310027, PR. China

[3]The University of Danang, University of Science and Education, 459 Ton Duc Thang, Lien Chieu, Danang, Vietnam

[4]School of Materials Science and Engineering, Harbin Institute of Technology, Harbin 150001, China

[5]School of Materials Science and Engineering, Inner Mongolia University of Technology, Hohhot 010051, P. R. China

[6]School of Natural Science, Ulsan National Institute of Science and Technology, Ulsan 44919, South Korea

A novel class of nanocrystalline/amorphous $Gd_3Ni/Gd_{65}Ni_{35}$ composite microwires was created directly by melt-extraction through controlled solidification. X-ray diffraction (XRD) and transmission electron microscopy (TEM) confirmed the formation of a biphase nanocrystalline/amorphous structure in these wires. Magnetic and magnetocaloric experiments indicate the large magnetic entropy change ($-\Delta S_M$ ~9.64 J/kg K) and the large refrigerant capacity (*RC* ~742.1 J/kg) around the Curie temperature of ~120 K for a field change of 5 T. These values are ~1.5 times larger relative to its bulk counterpart, and are superior to other candidate materials being considered for active magnetic refrigeration in the liquid nitrogen temperature range.

[*]Corresponding authors: hitshenhongxian@163.com (H.S.); faxiangqin@zju.edu.cn (F.X.Q.) phanm@usf.edu (M.H.P.)

## 1. Introduction

Magnetic materials exhibiting large magnetocaloric effects (MCEs) have been studied avidly due to its potential applications such as high efficiency cooling system without any detrimental to environment contrary to conventional gas compression refrigeration [1-7]. The principle of MCE results in a temperature change which can be achieved by varying applied magnetic field to the system with an adiabatic condition [1,2]. When it comes to cooling device, two factors including the magnetic entropy change ($\Delta S_M$) and the refrigerant capacity (*RC*) are considered to be important agents in a view point of the cooling efficiency, and the latter is closely related to magnetic transition mode [7]. The importance of *RC* and its relationship with a magnetic phase transition have been extensively studied [8-12], and the cooling efficiency is predominant when the material possesses a second-order magnetic transition (SOMT) rather than a first-order magnetic transition (FOMT), due to its negligible thermal and magnetic hysteresis losses [1-3, 13].

Specifically, rare earth elements such as Gd are regarded as promising magnetocaloric materials because they have SOMT character and relatively high $T_C$ which can be applied to room temperature working devices [2, 3, 6]. Until now many research groups have revealed that the shape and/or phase of sample effect on their MCE characteristics [14-21]. In particular, our group has recently demonstrated the excellent mechanical and magnetocaloric properties of amorphous Gd alloy microwires fabricated by a modified precision melt-extraction method [17, 18, 22-29]. Relative to their bulk counterparts, these microwires have shown larger MCE and *RC* values. The increased surface areas of the wires give rise to a higher heat transfer between the magnetic refrigerant and surrounding liquid [16]. Current efforts are to improve the MCE and *RC* of the microwires. In a recent study, we have demonstrated that the annealing process under the adequate atmosphere can improve its MCE characteristics also as releasing stress in the internal and surface of the material by supplying sufficient thermal energy during the process [22].

In this work, we demonstrate that controlled solidification of the melt-extraction process can crease novel biphase nanocrystalline/amorphous structures in Gd-based microwires for enhanced MCE and RC without performing thermal annealing. In particular, the large magnetic entropy change and refrigerant capacity are achieved in the nanocrystalline/amorphous $Gd_3Ni/Gd_{65}Ni_{35}$ composite microwires.

## 2. Experimental

The $Gd_3Ni/Gd_{65}Ni_{35}$ microwires were fabricated via the modified melt extraction method by using a spinning copper wheel of 160 mm in diameter with a $60^o$ tapered edge. To obtain uniform dimensions in the melt-extracted wires, the wheel was kept at a fixed rotation speed of 30m/min while the melted material was fed onto it at a constant rate of 30 μm/s. Energy dispersive X-ray spectroscopy (EDS) and scanning electron microscopy were performed with a JSM-6390LV model from JEOL and INCA x-sight by OXFORD Instruments. The microstructural evolution of the annealed wires was characterized by electron transmission microscopy (TEM). The magnetic property of a single wire with $d \sim 45\ \mu\text{m}$ and $l \sim 3\ \text{mm}$ was measured by a commercial physical property measurement system (PPMS) made by Quantum Design with a VSM option. DC magnetic fields of up to 5 T were applied longitudinally along the wire as the temperature was varied from 20 K to 180 K with an increment of 2 K.

## 3. Results and Discussion

### *3.1. Structural characterization*

The structural and morphological characterizations of melt-extracted $Gd_3Ni/Gd_{65}Ni_{35}$ microwires are shown in Fig. 1(a-c). The morphology and the diameter of the microwire were observed directly by planar and cross-sectional SEM micrographs, revealing a cylindrical wire of ~45 μm in diameter with a homogeneous surface (Fig. 1a). EDS analysis was performed to confirm the mixing ratio of the two elements quantitatively (Fig. 1b). The atomic percentages determined in this way are 65.95% and 34.05% for Gd and Ni, respectively. Both XRD (Fig. 1c) and TEM (inset of Fig. 1a) and SAED (inset of Fig. 1b) confirmed the presence of $Gd_3Ni$ nanocrystals of ~8 nm average diameter embedded in an amorphous matrix ($Gd_{65}Ni_{35}$). These two phases are expected to be magnetically coupled with each other, thus affecting the magnetic and magnetocaloric properties of the microwire.

### *3.2. Magnetic Properties*

The temperature dependence of magnetization $M(T)$ of the $Gd_3Ni/Gd_{65}Ni_{35}$ composite wire was measured in a field of 0.1T under a field-cooling protocol, and the result is shown in Fig. 2a. As one can see clearly in this figure, the microwire shows a broad paramagnetic to ferromagnetic (PM-FM) phase transition. This broadened transition can be attributed to the presence of two magnetic phases (the $Gd_3Ni$ nanocrystalline phase and the $Gd_{65}Ni_{35}$ amorphous phase) and the structural disorder caused by the amorphous structure [22]. It has been reported that crystalline (bulk) $Gd_3Ni$ exhibits an antiferromagnetic (AFM) ordering at $T_N$ ~99 K [30]. This $T_N$ is expected to shift to a lower temperature in the $Gd_3Ni/Gd_{65}Ni_{35}$ microwire due to the strong AFM weakening in the $Gd_3Ni$ phase caused by the nano-size effect. While the feature associated with the AFM ordering of the $Gd_3Ni$ phase is not obviously seen on the *M-T* curve (Fig. 2), we observe below a small, broad peak associated with it at ~50 K from the $-\Delta S_M(T)$ curves (Fig. 4). The Curie temperature ($T_C$) of the $Gd_3Ni/Gd_{65}Ni_{35}$ microwire determined from the minimum in *dM/dT* is ~120 K (see inset of Fig. 2a). This value of $T_C$ is close to that reported for an amorphous $Gd_{65}Ni_{35}$ ribbon ($T_C$ ~122 K) [31], but much larger than those reported for $Gd_{55}Ni_xAl_{45-x}$ ($x$ = 15, 20, 25, 30) bulk metallic alloys ($T_C$ ~ 50 – 70 K) [32].

To further understand the magnetic behavior of the $Gd_3Ni/Gd_{65}Ni_{35}$ microwire in the paramagnetic regime, the inverse susceptibility as a function of temperature, $\chi^{-1}(T) = \mu_o H/M$, generated from the $M(T)$ curve in the paramagnetic region is also shown in Fig. 2a. Its linear temperature dependence is consistent with the Curie – Weiss law in the paramagnetic region, $\chi = \frac{C}{T-\theta}$, where $C$ is the Curie constant defined as $C = \frac{N_A \mu_B^2}{3k_B}\mu_{eff}^2$ ($N_A$ = 6.022x10$^{23}$ mol$^{-1}$ is Avogadro number, $\mu_B$ = 9.274x10$^{-21}$ emu is the Bohr magneton, and $k_B$ = 1.38016x10$^{-16}$ erg/K is the Boltzmann constant). A linear fit of $\chi^{-1}(T)$ yields $\theta$ = 124 K and $C$ = 7.89 emu K mol$^{-1}$ T$^{-1}$. $\theta$ = 124 K is a positive value, which confirms the PM-FM phase transition. This Weiss temperature $\theta$ is very close to the Curie temperature $T_C$ (~120 K). This is likely associated with the presence of short-range magnetic order above the $T_C$.

The effective magnetic moment $\mu_{eff}$ of the sample is determined from $C$ via the following equation $\mu_{eff} = \sqrt{\frac{3k_B C}{N_A}} = \sqrt{8C}$. From this relationship, we obtained $\mu_{eff}$ = 7.941 $\mu_B$ for the presently fabricated $Gd_3Ni/Gd_{65}Ni_{35}$ microwires. This $\mu_{eff}$ value is approximately equal to the theoretically calculated value of pure Gd (~7.94 $\mu_B$). Some previous studies have reported this value of $\mu_{eff}$ for Gd-Ni [33-36]. Paulose *et al.* [33], Mallik *et al.* [34], and Uhlirova *et al.* [35] showed the parallel orientation of Gd and Ni moments. However, an antiferromagnetic coupling between the Gd and Ni moments was reported by Yano *et al.* [36]. In this work, the effective magnetic moment $\mu_{eff}$ is very close to pure Gd. That can be explained by the ferromagnetic coupling between Gd and Ni moments, or existence of non-negligible magnetic moments deriving from Ni-3d states.

Figure 2b presents the temperature dependence of magnetization at different applied magnetic fields, the $M(T)$ curves, from $\mu_0 H$ ~ 0.1 - 5 T. The PM-FM phase transition becomes broadened with increasing applied magnetic field up to 5 T, which is a typical behavior for SOMT materials. The largest changes in $M$ with respect to $T$ appear in a region from 120 K to 130 K, where $T_C$ has been calculated from the minimum of the d$M$/d$T$, indicating that the largest changes in magnetic entropy will occur in this temperature region [27].

To determine the nature of the PM-FM phase transition in the $Gd_3Ni/Gd_{65}Ni_{35}$ wire, the measured $M(H)$ isotherms for temperatures around the Curie temperature (Fig. 3a) have been converted into the Arrott plots ($\mu_0 H/M$ vs. $M^2$), as shown in Fig. 3b. According to the Banerjee criterion [37], a negative slope in a $\mu_0$H/M vs. $M^2$ plot indicates a discontinuity at the ferromagnetic transition. As can be seen from Fig. 3b, the slopes of the re-scaled isotherms are uniformly positive, thus the Banerjee criterion for a second-order magnetic transition (SOMT) is satisfied in this system. This result is in full agreement with our previous reports on Gd-based microwires [17-19].

*3.3. Magnetocaloric effect*

The magnetic entropy change in the system due to a field change between 0 and $H_{max}$ is calculated using the thermodynamic Maxwell relation:

$$\Delta S_M(T, \mu_o H) = \mu_0 \int_0^{H_{max}} \left(\frac{\partial M}{\partial T}\right)_H dH \quad (1)$$

where $M$ is the magnetization, $\mu_o H$ is the applied magnetic field, and $T$ is the temperature, by integrating over the magnetic field. The resulting $-\Delta S_M(T)$ curves for several applied field changes are shown for the $Gd_3Ni/Gd_{65}Ni_{35}$ microwire in Fig. 4a,b. In each case, a broad peak occurs centered around the $T_C$, reaching a large maximum value of $|\Delta S_M| = 9.64$ J/kg K for an applied field change of 5 T. This value is 1.4 times higher than that found in the ribbon sample of the $Gd_{65}Ni_{35}$ composition [31]. It is worth noting that in addition to the dominant peak of the $-\Delta S_M(T)$ curves associated with the PM-FM phase transition at $T_C$ ~120 K, a small, broad peak around ~50 K is also observed and attributed to the AFM transition of the nanocrystalline $Gd_3Ni$ phase. Since this AFM phase is considerably weak at the nanoscale, application of a sufficiently high magnetic field may convert it into FM, giving rise to the broad $-\Delta S_M(T)$ curve and hence the large refrigerant capacity (*RC*) - an important figure of merit in magnetic cooling.

The $RC$ of the $Gd_3Ni/Gd_{65}Ni_{35}$ microwire has been calculated from the data by integrating the area under the $-\Delta S_M$-$T$ curves by using the temperatures at half maximum of the peaks:

$$\mathrm{RC} = \int_{\mathrm{T_{hot}}}^{\mathrm{T_{cold}}} -\Delta S_{\mathrm{M}}(T)\mathrm{d}T \qquad (2)$$

where $T_{cold}$ and $T_{hot}$ are the onset and offset temperatures of $\delta T_{FWHM}$, respectively. The other method, using the relative cooling power (RCP), represents an amount of heat transfer between the hot and cold sides in an ideal refrigeration cycle, which can be defined as Wood and Potter's method:

$$\mathrm{RCP} = -\Delta S_{\mathrm{M}}^{\mathrm{max}}\, \delta T_{\mathrm{FWHM}} \qquad (3)$$

where $\delta T_{FWHM} = T_{hot} - T_{cold}$ is the temperature difference at the full width at half maximum of the $\Delta S_M$ ($T$) curve. Values of RC and RCP are plotted as functions of magnetic field, as shown in Fig. 5a. For comparison, the RC and RCP values of the present microwire and other candidate materials for a field change of 5 T are summarized in Table 1 and selectively plotted in Fig. 5b. The $RC$ of ~742 J/kg of the $Gd_3Ni/Gd_{65}Ni_{35}$ microwire is enhanced over bulk $Gd_{65}Ni_{35}$ (~524 J/kg), and is also 40% larger than the $RC$ of the ribbon with the same composition [31]. It can be seen that the $Gd_3Ni/Gd_{65}Ni_{35}$ microwire shows almost the largest $RC$ among the compared candidates. While the RC is almost equal for both $Gd_{53}Al_{24}Co_{20}Zr_3$ and $Gd_3Ni/Gd_{65}Ni_{35}$ microwires, the $Gd_3Ni/Gd_{65}Ni_{35}$ microwire has a significantly higher $T_C$ which is more desirable for magnetic refrigeration in the liquid nitrogen temperature regime. In general, magnetocaloric microwires with AM/AM+NC structures represent significantly higher RC and RCP values than RG/BMG structures [22,23,28].

## 4. Conclusion

We have shown that a novel biphase nanocrystalline/amorphous structure can be created in Gd alloy microwires such as $Gd_3Ni/Gd_{65}Ni_{35}$ microwires directly from melt-extraction through controlled solidification. These microwires possess enhanced Curie temperature and RC as compared to their bulk and ribbon counterparts. The excellent magnetocaloric properties make the composite $Gd_3Ni/Gd_{65}Ni_{35}$ microwires an attractive candidate material for applications as a cooling device for micro electromechanical systems and nano electro mechanical systems.

**Acknowledgments**

Work at USF was supported by the U.S. Department of Energy, Office of Basic Energy Sciences, Division of Materials Sciences and Engineering under Award No. DE-FG02-07ER 46438 (Magnetocaloric studies). Work at HIT was supported by the National Natural Science Foundation of China (NSFC, Nos. 51671071) (Sample fabrication and TEM characterization). Work at ZJU was supported by the National Natural Science Foundation of China (NSFC, Nos. 51671171). The research work at Korea was supported by the National Research Foundation of Korea under Grant No: 2020R1A2C1008115 (Magnetic measurements).

## Figure Captions

**Figure 1.** (a) SEM image and inset of (a) shows an TEM morphology; (b) Energy dispersive spectrometry (EDS) of the local region in the TEM image and inset of (b) shows a selected area electronic diffraction (SAED) of the nanocrystalline microstructure including $Gd_3Ni$ phase; (c) X-ray diffraction pattern (XRD) of the $Gd_3Ni/Gd_{65}Ni_{35}$ microwires.

**Figure 2.** (a) Temperature dependence of magnetization $M(T)$ and its derivative (dM/dT) taken in a field of 0.1 T for the $Gd_3Ni/Gd_{65}Ni_{35}$ wire. (b) The filled 2D contour plot of the temperature and field dependence of the magnetization.

**Figure 3.** (a) Magnetic field dependence of magnetization M(H) taken at temperatures around the Curie temperature. (b) Arrott-type plots ($\mu_0 H/M$ vs. $M^2$) of the $Gd_3Ni/Gd_{65}Ni_{35}$ wire.

**Figure 4.** (a) Temperature dependence of magnetic entropy change $-\Delta S_M(T)$ of the $Gd_3Ni/Gd_{65}Ni_{35}$ microwire for selected magnetic field changes. (b) The filled 2D contour plot of the temperature and field dependence of the magnetic entropy change.

**Figure 5.** Refrigerant capacities (*RC*) of various magnetocaloric candidate materials in the temperature range of interest 10 – 200 K. Red indicates the microwires, while Blue indicates their bulk counterparts for one-to-one comparison.

**Figure 1**

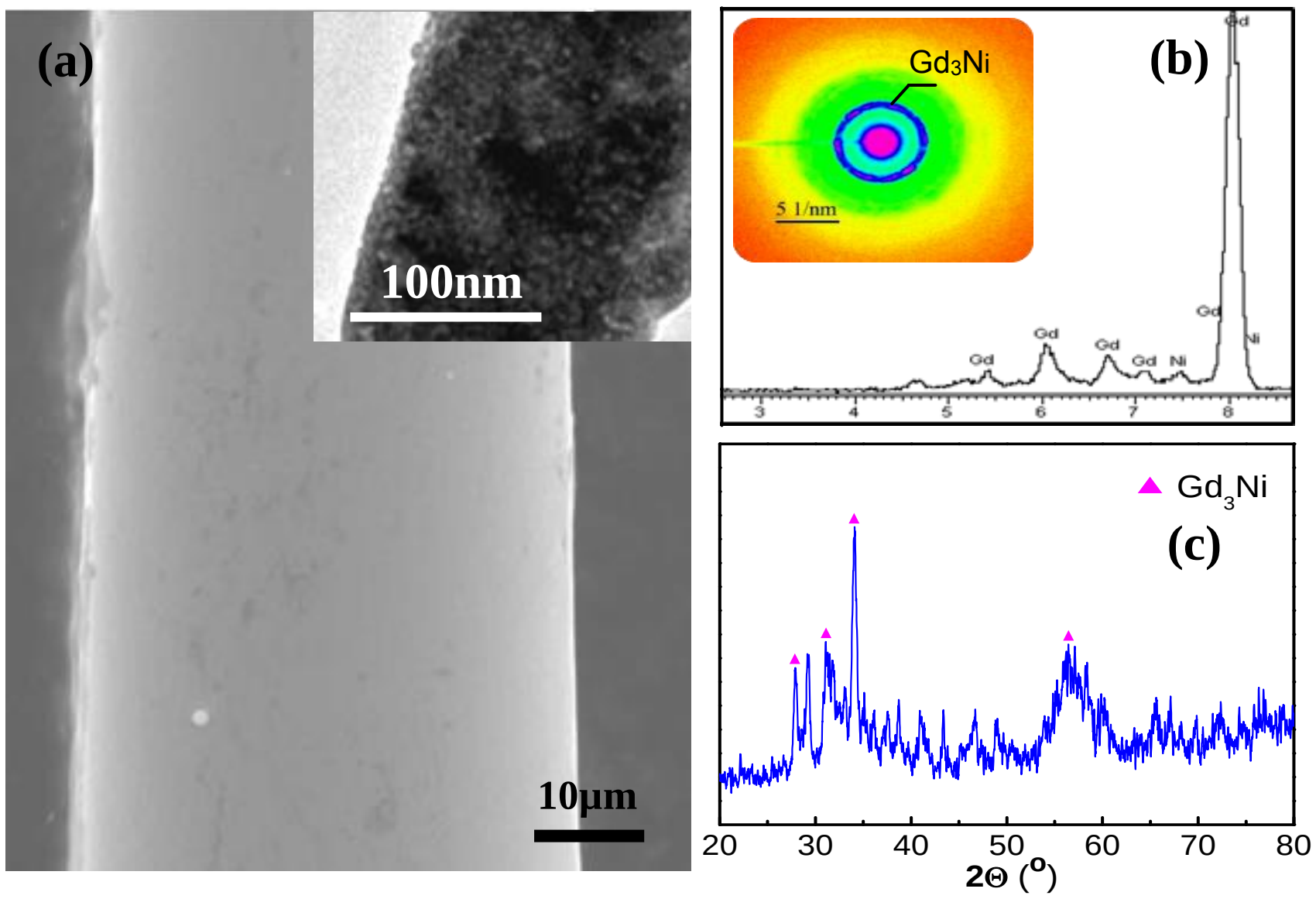

**Figure 2**

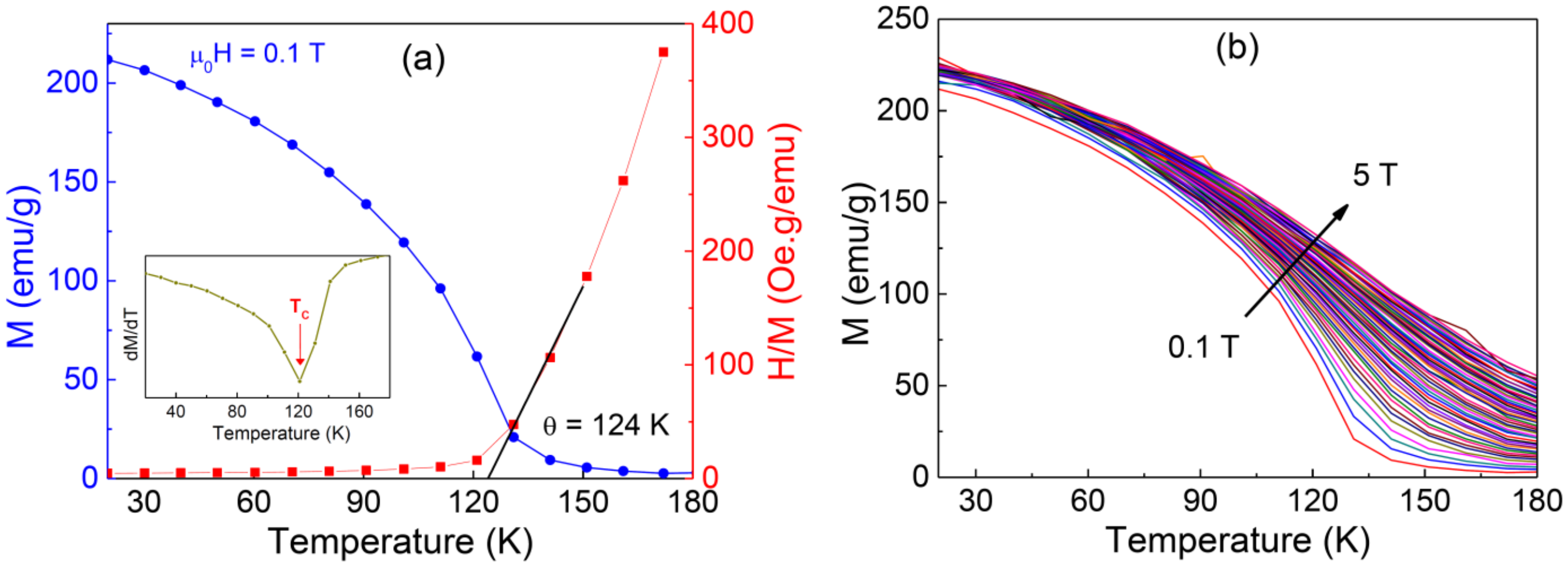

**Figure 3**

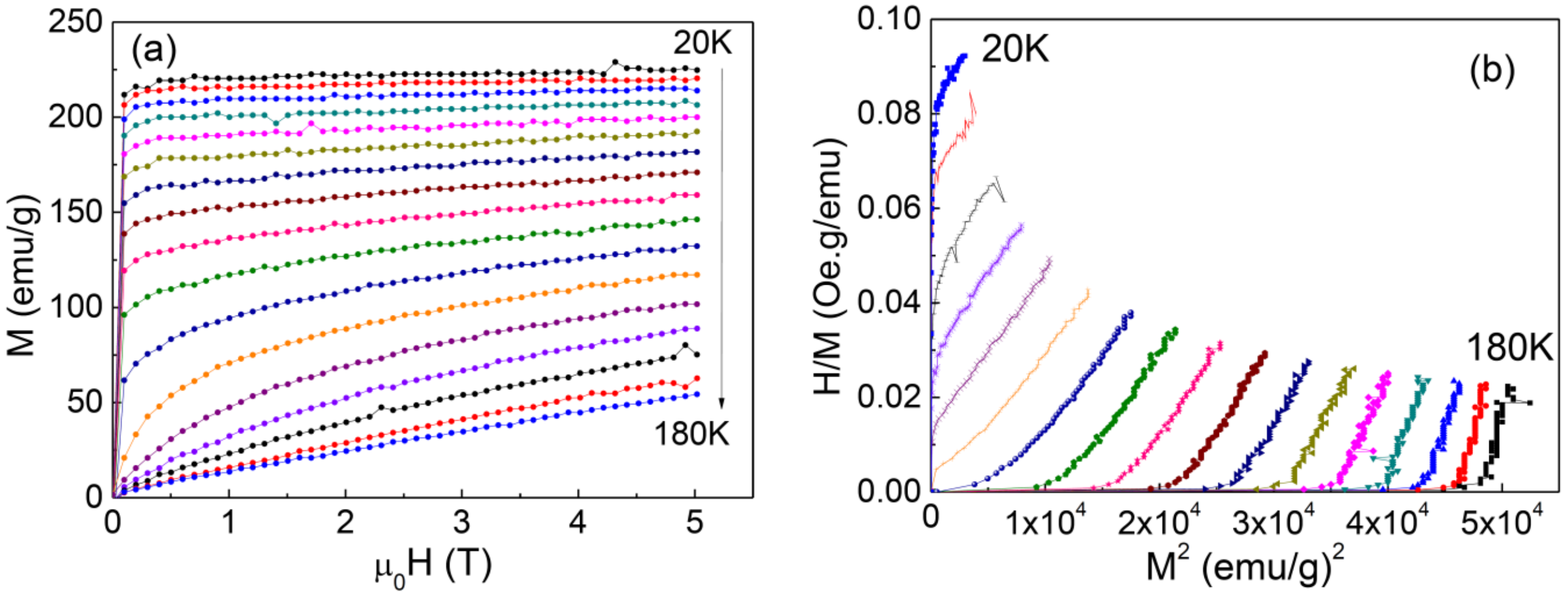

**Figure 4**

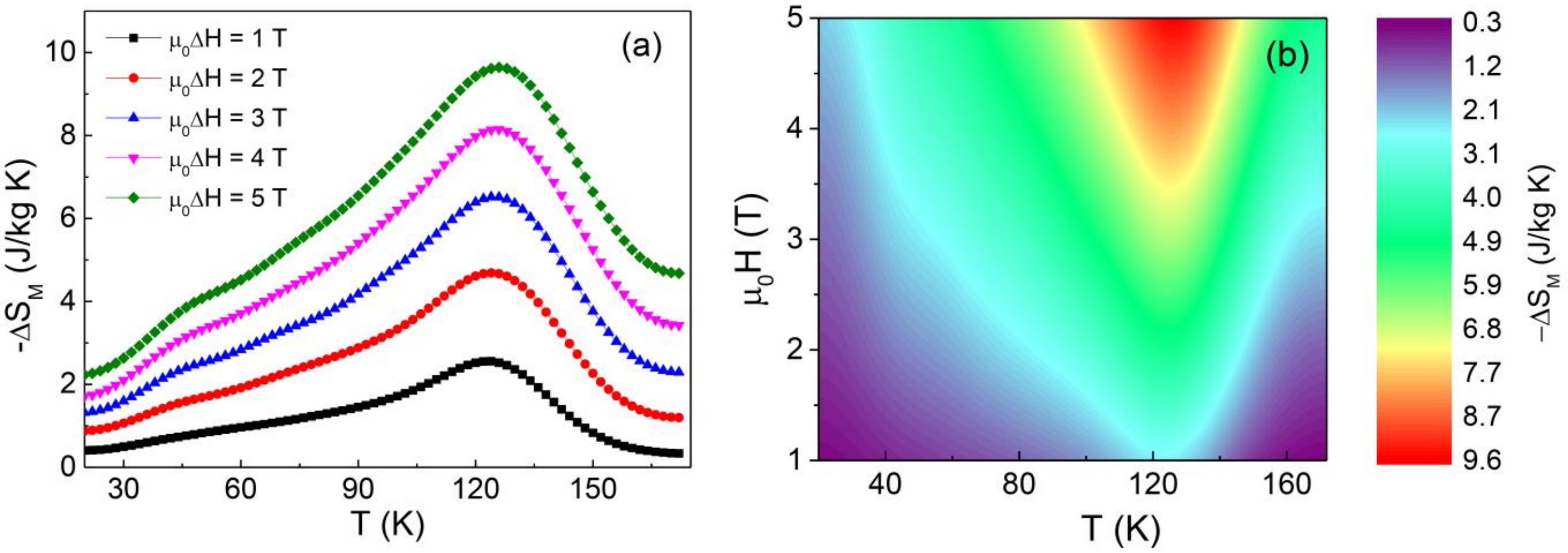
(a)
$\mu_0\Delta H$ = 1 T
$\mu_0\Delta H$ = 2 T
$\mu_0\Delta H$ = 3 T
$\mu_0\Delta H$ = 4 T
$\mu_0\Delta H$ = 5 T
$-\Delta S_M$ (J/kg K)
T (K)
(b)
$\mu_0H$ (T)
T (K)
$-\Delta S_M$ (J/kg K)
0.3
1.2
2.1
3.1
4.0
4.9
5.9
6.8
7.7
8.7
9.6

**Figure 5**

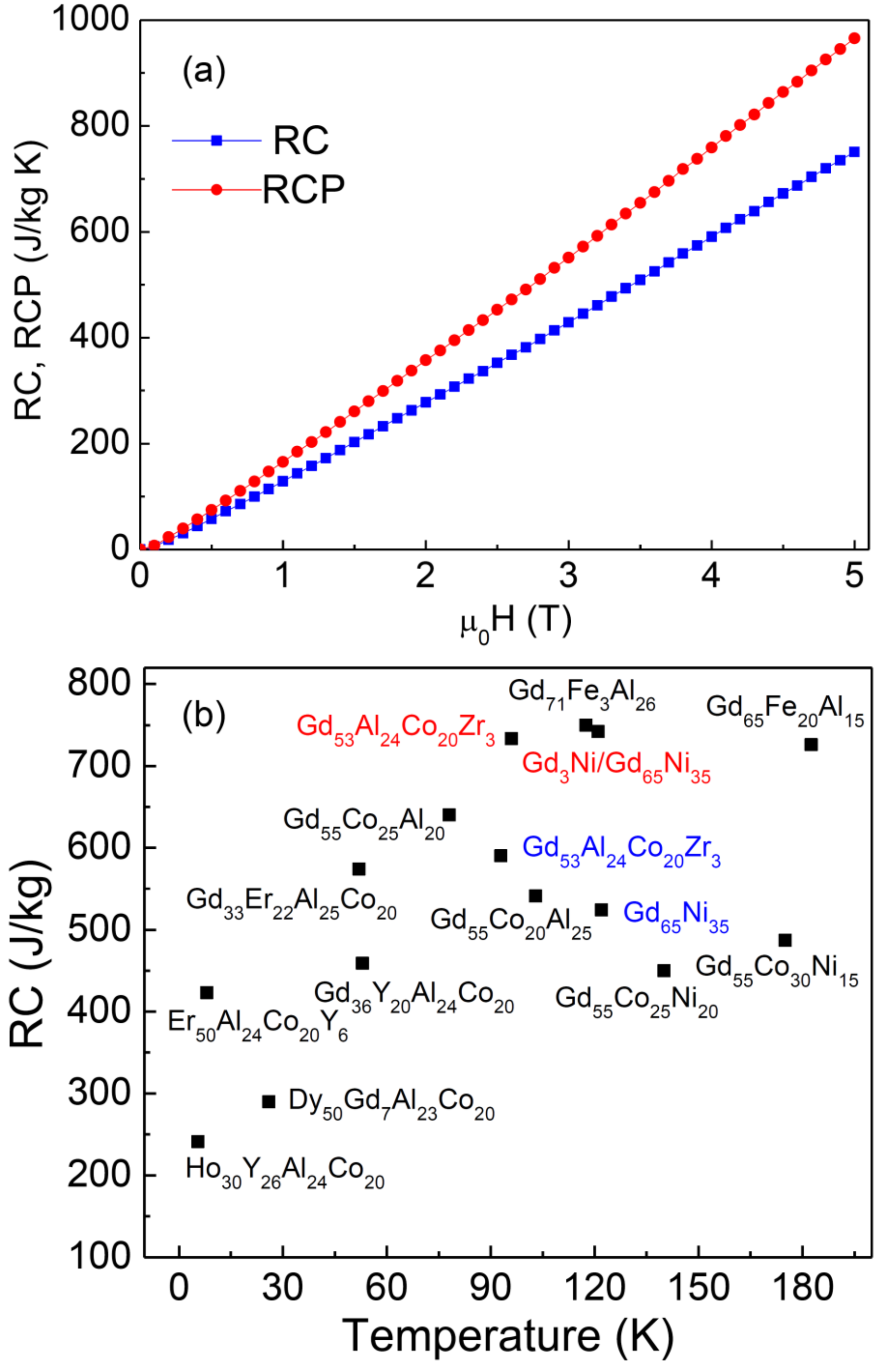

**Table 1.** Comparison of the magnetocaloric characteristics of the present $Gd_{65}Ni_{35}$ microwires and other MCE materials depending on their phase structures.

| Material | Structure | *Tc* (K) | $-\Delta S_M$ (J/kg K) | RC (J/kg) | RCP (J/kg) | Applied Field (T) | Reference |
|---|---|---|---|---|---|---|---|
| $Gd_{65}Ni_{35}$ | AM+NC | 121 | 9.64 | 742.1 | 965.3 | 5 | This work |
| $Gd_{65}Ni_{35}$ | GR | 122 | 6.9 | 524 | - | 5 | [38] |
| $Gd_{73.5}Si_{13}B_{13.5}/GdB_6$ | AM+NC | 106 | 6.4 | 790 | 885 | 5 | [28] |
| $Gd_{50}Al_{30}Co_{20}$ | AM+NC | 86 | 10.09 | 672 | 861 | 5 | [23] |
| $Gd_{55}Al_{25}Co_{20}$ | AM+NC | 100 | 9.67 | 652 | 861 | 5 | [39] |
| $Gd_{60}Al_{20}Co_{20}$ | AM+NC | 109 | 10.11 | 681 | 915 | 5 | [40] |
| $Gd_{53}Al_{24}Co_{20}Zr_3$ (Annealed at 100 $^oC$) | AM+NC | 94 | 9.5 | 687 | 893 | 5 | [22] |
| $Gd_{53}Al_{24}Co_{20}Zr_3$ (Annealed at 200 $^oC$) | AM+NC | 94 | 8.0 | 629 | 744 | 5 | [22] |
| $Gd_{53}Al_{24}Co_{20}Zr_3$ (Annealed at 300 $^oC$) | AM+NC | 94 | 5.1 | 396 | 525 | 5 | [22] |
| $Gd_{60}Al_{25}Co_{15}$ | AM | 101 | 9.73 | 732 | 973 | 5 | [41] |
| $Gd_{55}Al_{20}Co_{25}$ | AM | 110 | 9.69 | 580 | 804 | 5 | [42] |
| $Gd_{50}$-$(Co_{69.25}Fe_{4.25}Si_{13}B_{13.5})_{50}$ | AM | 170 | 6.56 | 625 | 826 | 5 | [25] |
| $Gd_{50}$-$(Co_{69.25}Fe_{4.25}Si_{13}B_{13.5})_{50}$ | AM | 174 | 5.90 | - | - | 5 | [12] |
| $Gd_{60}Fe_{20}Al_{20}$ | AM | 202 | 4.8 | 687 | 900 | 5 | [27] |
| $Gd_{59.4}Al_{19.8}Co_{19.8}Fe_1$ | AW | 113 | 10.33 | 748.22 | 1006 | 5 | [43] |
| $Gd_{53}Al_{24}Co_{20}Zr_3$ | MW | 96 | 10.3 | 733.4 | - | 5 | [17] |
| $Gd_{55}Co_{25}Ni_{20}$ | GR | 140 | 6.04 | 450 | - | 5 | [44] |
| $Gd_{55}Co_{30}Ni_{15}$ | GR | 175 | 6.3 | 487 | - | 5 | [44] |
| $Gd_{55}Co_{20}Al_{25}$ | BMG | 103 | 8.8 | 541 | - | 5 | [45] |
| $Gd_{55}Co_{25}Al_{20}$ | BMG | 78 | 8 | 640 | - | 5 | [45] |
| $Gd_{53}Al_{24}Co_{20}Zr_3$ | BMG | 93 | 9.4 | 590 | - | 5 | [46] |
| $Gd_{33}Er_{22}Al_{25}Co_{20}$ | BMG | 52 | 9.47 | 574 | - | 5 | [46] |
| $Ho_{30}Y_{26}Al_{24}Co_{20}$ | BMG | 5.5 | 10.76 | 241 | - | 5 | [47] |
| $Dy_{50}Gd_7Al_{23}Co_{20}$ | BMG | 26 | 9.77 | 290 | - | 5 | [47] |
| $Er_{50}Al_{24}Co_{20}Y_6$ | BMG | 8 | 15.91 | 423 | - | 5 | [47] |
| $Gd_{71}Fe_3Al_{26}$ | BMG | 117.5 | 7.4 | 750 | - | 5 | [48] |
| $Gd_{65}Fe_{20}Al_{15}$ | BMG | 182.5 | 5.8 | 726 | - | 5 | [48] |
| $Gd_{36}Y_{20}Al_{24}Co_{20}$ | BMG | 53 | 7.76 | 459 | - | 5 | [49] |

AM: Amorphous Microwires, NC: Nanocrystals, GR: Glassy Ribbon, BMG: Bulk Metallic Glass